# Neurodiversity and Technostress: Towards a Multimodal Research Design for Evaluating Subjective, Physiological, and Behavioral Responses

Lisa van den Heuvel[1](✉), Igor Ivkić[1,2], René Riedl[3,4]

[1]*University of Applied Science Burgenland, Eisenstadt, AT*
[2]*Lancaster University, Lancaster, UK*
[3]*University of Applied Sciences Upper Austria, Steyr, AT*
[4]*Johannes Kepler University, Linz, AT*

*2410781017@hochschule-burgenland.at*, *i.ivkic@lancaster.ac.uk, rene.riedl@fh-steyr.at*

**Abstract.** Digitalization has transformed modern work by increasing efficiency while also introducing new forms of strain. Technostress (TS) describes subjective, physiological, and behavioral stress responses related to digital technology use. Existing TS research has predominantly focused on neurotypical populations and rarely integrates multiple stress dimensions within a single design. This paper addresses these gaps by proposing a controlled experimental research design that systematically compares neurodivergent and neurotypical individuals under standardized digital stress conditions. The proposed design combines structured and unstructured digital tasks with a multimodal measurement approach covering subjective perceptions, physiological activation, and observable interaction behavior. By integrating neurodiversity into TS research, the paper contributes to a more differentiated understanding of digital stress and provides a methodological approach for more inclusive digital work design.

## 1 Introduction

Digital technologies have become a central part of modern work environments. Information and communication technologies enable new forms of collaboration, increase flexibility, and structure core work processes [1, 2, 3]. At the same time, the growing reliance on digital systems is associated with increasing experiences of overload, interruption, and expectations of constant availability. These phenomena are commonly summarized under the term Technostress (TS), also referred to as digital stress [1, 2, 4, 5].

TS describes stress reactions that arise in response to the use of digital technologies, particularly when these technologies are experienced as complex, interruptive, or difficult to control [1, 2]. Prior research has shown that TS can negatively affect well-being, job satisfaction, and performance, and may also influence work behavior and organizational outcomes [2, 4-6]. At the same time, recent studies emphasize that digital technologies can also act as supportive resources, for example by enabling autonomy or improving work processes, highlighting the dual nature of technology as both a stressor and a resource [6-8].

This dual role highlights that TS should not be understood as a purely negative phenomenon, but rather as the result of an interaction between technological demands, available resources, and individual characteristics [8]. In this context, digital stress depends not only on technological features but also on how individuals perceive and process these demands. However, existing research has primarily focused on neurotypical (NT) populations, largely overlooking differences in cognitive and sensory processing across individuals.

Open questions remain regarding how different groups of individuals perceive and process digital stressors. In particular, neurodivergent (ND) individuals, such as those with attention-deficit/hyperactivity disorder (ADHD), autism spectrum conditions, or dyslexia, often show distinct patterns in attention regulation, executive functioning, and sensory processing [10-13]. As these characteristics are highly relevant in digital work environments, ND individuals may experience and process digital stressors differently compared to NT individuals.

Despite the increasing relevance of neurodiversity in the workplace, empirical research that systematically examines these differences in the context of TS remains limited [1, 2, 4]. Furthermore, existing TS research often focuses on either subjective, physiological, or behavioral indicators, while multimodal approaches have only rarely been applied to compare different user groups within a single research design [15-19]. This constitutes an important limitation, as stress is a multi-layered phenomenon in which subjective perception, physiological activation, and behavioral responses may diverge [15].

To overcome this limitation, this paper proposes a controlled experimental research design that systematically compares ND and NT individuals under standardized digital stress conditions. By integrating subjective, physiological, and behavioral measures within a multimodal framework, the paper aims to contribute to a more differentiated understanding of TS. This design enables the analysis of how ND and NT individuals respond to digital stressors and provides a basis for designing more inclusive digital work environments.

## 2     Background and Research Gap

This section summarizes key concepts related to TS and neurodivergence, identifies the research gap addressed in this paper, reviews TS in digital workplaces, examines how neurodivergence influences information processing and stress responses, and highlights the need for a multimodal research approach.

## 2.1 Technostress in Digital Work Contexts

Research on TS primarily originates from the fields of information systems (IS) and occupational psychology. Previous studies have identified core categories of technology-related stressors, including techno-overload, techno-invasion, techno-complexity, and techno-uncertainty [1, 2, 4]. These stressors describe situations in which individuals experience excessive demands, blurred work–life boundaries, difficulties in using technologies, or constant technological change.

More recent research suggests that traditional TS frameworks may not fully capture the evolving nature of digital work environments. Emerging stressors such as system unreliability or increased monitoring, or constant tool switching have been identified as additional sources of strain [8, 20]. In response, newer measurement approaches, such as the digital stressors scale (DSS), conceptualize TS as a multidimensional construct that reflects a broader range of technology-related demands [21].

TS has been associated with a variety of negative outcomes, including cognitive overload, reduced job satisfaction, decreased performance, and health-related impairments [1, 2, 4, 6]. It has also been linked to negative organizational outcomes, such as changes in work behavior and increased tendencies toward knowledge hiding [22]. Experimental research further indicates that digital stressors are associated with physiological stress responses, including changes in heart rate variability (HRV), blood pressure, and cortisol levels [15, 17, 18, 23, 24]. These physiological responses reflect activation of the autonomic nervous system and provide objective indicators of stress.

At the same time, digital technologies should not be viewed exclusively as sources of stress but can also function as supportive resources. Depending on their design, the context of use, and individual characteristics, digital technologies can be experienced as a supportive resource, for example, when they enhance autonomy or facilitate work processes [6, 7]. Concepts such as techno-eustress describe situations in which technology use is experienced as motivating or performance-enhancing [6], while techno-relief refers to the potential of digital tools to reduce stress and support work processes [25]. In this context, digital technologies can act both as hindrance stressors (e.g., interruptions, overload) and as challenge stressors (e.g., stimulating tasks, autonomy) [26].

Given the multidimensional and context-dependent nature of TS, researchers have emphasized the importance of integrating subjective, physiological, and behavioral indicators, as these may capture different but complementary aspects of stress [16, 27]. However, and to the best of our knowledge, such multimodal approaches have rarely been applied in a comparative context across different user groups.

## 2.2 Neurodivergence and Stress Processing

Alongside TS research, the concept of neurodiversity has gained increasing attention in organizational and information systems contexts. Neurodiversity refers to natural variations in how individuals perceive, process, and respond to information, including conditions such as ADHD, autism spectrum conditions, and dyslexia [10–13]. Recent IS research also argues that neurodiversity should be considered more systematically because neurodiversity facets can influence cognition, emotion, decision-making, and user behavior, and because ignoring such differences may reduce the inclusivity and explanatory power of IS research [28].

While neurodiversity encompasses a broad spectrum of conditions, this study focuses on selected forms of neurodivergence that are particularly relevant in digital work environments. Specifically, ADHD, autism spectrum conditions, and dyslexia are considered, as they are associated with differences in attention regulation, executive functioning, and sensory processing that are highly relevant in technology-mediated work settings [29].

Prior research suggests that individuals with ADHD may experience difficulties in sustained attention, impulse control, and executive functioning, which can affect their ability to manage multiple tasks, maintain focus, and resist distractions in digital environments [14]. Autism spectrum conditions are associated with differences in sensory processing and information filtering, which may increase sensitivity to environmental stimuli and contribute to sensory overload in complex or dynamic digital interfaces [14, 30]. Dyslexia, characterized by persistent difficulties in reading and processing written language, may influence how individuals interact with text-intensive digital systems and increase cognitive effort in information processing tasks [14]

These characteristics are highly relevant in digital work contexts and may lead to differences in subjective, physiological, and behavioral stress responses. For example, differences in attention regulation may be reflected in interaction patterns such as task-switching frequency, response times, or error rates, while sensory sensitivity may influence visual attention and gaze behavior in complex interfaces [18].

At the same time, workplace research highlights that ND individuals can contribute specific strengths, such as attention to detail, pattern recognition, or problem-solving abilities, particularly in supportive environments [10-13]. However, despite these insights, quantitative and experimental research that directly compares ND and NT individuals in controlled digital work scenarios remains limited.

### 2.3 Research Gap and Research Questions

While prior research has provided valuable insights into TS and, separately, into neurodiversity in the workplace, these research streams have largely evolved independently. TS research has predominantly focused on NT populations, whereas neurodiversity research has mainly examined cognitive characteristics, inclusion, and workplace strengths and challenges. As a result, there is limited empirical evidence on how ND and NT individuals differ in their perception of and response to digital stressors in work-related contexts.

In addition, TS is increasingly conceptualized as a multidimensional and context-dependent phenomenon. However, many studies focus on isolated aspects, such as self-reported perceptions or individual physiological indicators. Although multimodal approaches have been proposed in prior research, they have rarely been applied to systematically compare ND and NT individuals in controlled digital work environments [27]. This is particularly relevant, as subjective perception, physiological activation, and observable behavior may diverge.

Furthermore, existing research is predominantly based on survey-based or cross-sectional designs, while controlled experimental studies in realistic digital work scenarios remain limited. This constrains the ability to examine causal relationships and to compare how different user groups respond to controlled digital stress conditions.

Against this background, this study proposes a controlled experimental research design that systematically compares ND and NT individuals under controlled digital stress conditions. The design integrates subjective perceptions, physiological stress responses, and observable interaction behavior in a multimodal approach.

Based on this approach, the following research questions (RQ) are proposed:

- **RQ1:** To what extent do ND and NT individuals differ in their subjective perception of digital stressors in a controlled digital work scenario?
- **RQ2:** To what extent do ND and NT individuals differ in their physiological stress reactions (heart rate, heart rate variability) under digital stress conditions?
- **RQ3:** To what extent do ND and NT individuals differ in observable work and interaction behavior under digital stress conditions?

## 3   Methods

To answer the proposed research questions, this section presents the methodological framework of the study. It describes the experimental design, participant selection, procedure, measurement approach, and planned data analysis. The aim is to provide a coherent and transparent description of how differences in TS responses between ND and NT individuals can be examined in a controlled setting.

### 3.1   Research Design

This paper proposes a controlled experimental research design to systematically investigate how ND and NT individuals perceive and respond to digital stressors. To capture group-specific and context-dependent effects, the design combines a between-subjects factor and a within-subjects factor. Neurodivergence (ND and NT groups) serves as the between-subjects factor, while participants complete different digital task types as within-subject conditions.

This approach allows comparison of stress responses between groups, while also examining how the same individuals react across different digital work contexts. In addition, the use of two task types supports a more differentiated interpretation of results and improves the potential generalizability of the proposed design.

An overview of the proposed experimental flow is provided in Figure 1. The procedure consists of a baseline phase, a task phase with counterbalanced task order, an immediate post-task phase, and a follow-up assessment capturing perceptions of digital stress in everyday work settings.

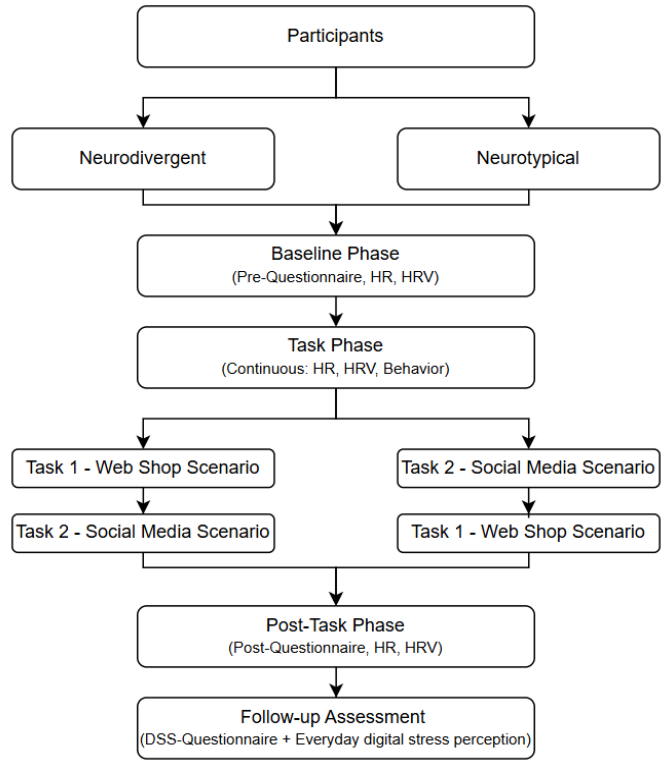

**Fig. 1.** Overview of the proposed experimental design

### 3.2 Participants

As shown in Fig. 1, the proposed study aims to include both ND and NT participants. ND status would be operationalized through self-report and would focus on forms of neurodivergence that are particularly relevant to digital work environments, including ADHD, autism spectrum conditions, and dyslexia. Because the study is concerned with work-related digital stress, the target sample should consist of adults who regularly use digital tools in work-related or closely comparable settings.

Participants could be recruited through online channels, professional networks, and, where appropriate, university populations with relevant digital work experience. In addition to group membership, demographic and contextual information, such as age, gender, education and prior experience with digital tools, would be collected to account for potential confounding influences and ensure comparability between groups [31].

With regard to sample size, the proposed design should aim for at least 50 participants per group as a lower bound for meaningful group comparisons. Larger samples would be preferable to increase statistical power but may not be feasible within practical constraints. Because this paper presents a research design rather than an implemented study, this number should be understood as a design-oriented benchmark rather than as a finalized recruitment target.

## 3.3 Experimental Tasks

To approximate realistic digital work environments, the proposed design includes two experimental tasks reflecting different forms of digital interaction and stress exposure:

**Task 1: Structured task – Web shop scenario.** The first task is a structured, goal-oriented scenario based on a web shop environment. Participants are asked to complete a clearly defined task, such as placing an online order. During task execution, controlled digital stressors are introduced, including interruptions, changing requirements, system delays, temporary malfunctions or crashes, and time pressure. This task is intended to represent process-driven digital work with explicit goals and external constraints.

**Task 2: Unstructured task – Social media scenario.** The second task is a more unstructured and dynamic scenario based on a social-media-like environment. Participants perform loosely defined activities, such as interacting with content, writing short responses, maintaining a chat conversation, and switching between parallel activities. In contrast to the web shop task, this scenario is intended to reflect less structured digital environments characterized by continuous information flow, multitasking, and higher demands on cognitive flexibility.

Across both tasks, TS is induced through experimentally controlled stressors embedded in the digital environment, including interruptions, multitasking demands, time pressure, and system-related issues. This approach allows digital stress to be represented in a controlled yet ecologically plausible manner.

To reduce potential order effects, the sequence of tasks would be systematically varied across participants. One subgroup would start with the structured task and then complete the unstructured task, while the other subgroup would follow the reverse order. This counterbalancing procedure is intended to minimize order effects such as fatigue, learning, task familiarity, and potential carry-over effects between tasks.

## 3.4 Procedure

The following phases structure the experimental procedure and capture stress responses across baseline, task execution, post-task assessment, and follow-up self-report. To reduce expectation effects, participants are not explicitly informed that stress represents the main focus of the study. Instead, a neutral description of the experiment is provided to minimize anticipatory stress and potential response biases.

**Baseline Measurements Phase.** The first phase would consist of a short baseline period under resting conditions. During this phase, heart rate (HR) and HRV would be recorded to establish individual physiological reference values. In addition, participants would complete a brief pre-task questionnaire assessing their current condition, including perceived energy level, sleep quality, recent physical activity, caffeine or nicotine intake, and medication use. This information is intended to document factors that may influence physiological stress responses independently of the experimental tasks.

**Task Phase.** In the second phase, participants would complete both experimental tasks in counterbalanced order. During task execution, physiological and behavioral data would be recorded continuously. This setup would allow stress responses to be examined in relation to specific task events, such as interruptions, delays, or system errors.

**Post-Task Phase.** The third phase would consist of an immediate post-task assessment. Physiological recording would continue for a short period after task completion in order to capture short-term recovery processes. Participants would also complete a post-task questionnaire assessing perceived stress, task difficulty, and subjective experience during the tasks. Following task completion, participants would be debriefed about the purpose of the experiment and given the opportunity to ask questions.

**Follow-up Assessment.** In addition, the design proposes a follow-up self-report assessment after a defined interval. This follow-up would include standardized TS-related instruments, such as the DSS [21], as well as semi-structured questions on how participants perceive digital stress in their everyday work contexts. The purpose of this follow-up is to complement the immediate laboratory-based assessment with a less situation-specific perspective on digital stress in daily work life.

### 3.5 Measures

To reflect the multidimensional nature of TS, the proposed design follows a multimodal measurement approach that integrates the following subjective, physiological, and behavioral indicators:

**Subjective stress.** Subjective stress would be assessed at several points in time. Pre-task measures would capture the participants' initial condition before stress induction. Immediate post-task measures would assess perceived stress, task difficulty, and subjective experience during the experimental tasks. In addition, the follow-up assessment would be used to capture more stable and context-related perceptions of digital stress in everyday work. For this assessment, standardized instruments such as the DSS [21] could be combined with Likert-scale items and selected semi-structured questions to capture subjective experiences more comprehensively.

**Physiological measures.** Physiological measures would include HR and HRV, recorded using a wearable chest-strap device. Measurements would be taken during baseline, throughout task execution, and during the immediate post-task phase. Baseline values are important because they provide an individual physiological reference point and allow stress-related changes to be interpreted relative to each participant's resting state. Physiological responses reflect activation of the autonomic nervous system and provide objective indicators of stress. In particular, HRV-based metrics such as the root mean square of successive differences (RMSSD) could be analyzed to capture short-term variations in autonomic regulation, especially parasympathetic activity [32].

**Behavioral data.** Behavioral data are collected during task execution using interaction-tracking methods, including mouse movements, click behavior, response times, and task-switching patterns. These measures help capture observable responses to digital stressors, such as increased hesitation, error-proneness, or inefficient interaction patterns. If feasible, eye-tracking can be integrated to assess visual attention and cognitive processing during task performance. Metrics such as fixation duration and gaze transitions provide additional insights into attentional strategies and cognitive load in complex digital environments.

### 3.6 Data Analysis

The proposed data analysis follows the multimodal research design and integrates subjective, physiological, and behavioral indicators. Group comparisons between ND and NT can be conducted using statistical methods appropriate for the mixed design, such as repeated measures analysis of variance (ANOVA), to assess within-subject effects across task types, between-group differences, and interaction effects between group membership and task condition.

Physiological data can be analyzed relative to baseline values to assess changes in activation during the task phase and short-term recovery in the post-task phase. Depending on data structure and quality, mixed-effects models may be considered, particularly in the presence of repeated or unbalanced observations.

In addition, correlation and regression analyses can examine relationships between subjective stress perception, physiological responses, and behavioral indicators, allowing the evaluation of convergence or divergence between perceived and objectively measured stress responses. Such discrepancies are especially relevant in TS, where physiological activation, subjective appraisal, and behavior may not always align.

Finally, exploratory analyses can consider whether contextual or individual variables, such as prior experience with digital tools, are associated with stronger or weaker stress responses. These analyses are exploratory rather than confirmatory and can serve as a basis for future research.

### 3.7 Ethical Considerations

The proposed study induces and measures TS in a controlled laboratory setting and therefore requires clear ethical safeguards to protect participants' well-being and autonomy. Participants would receive information on the procedure and measurement methods, provide written informed consent, and be free to withdraw at any time without disadvantage.

Although the tasks are designed to induce digital stress, the procedure would remain within the range of typical workplace-related strain rather than excessive burden. Participants would be monitored throughout the experiment, and the procedure would be stopped if signs of significant distress occur. After the study, participants would be debriefed and given the opportunity to discuss their experience. All collected data would be stored pseudonymously and used solely for research purposes in compliance with data protection regulations.

## 4      Discussion and Outlook

This paper addresses the underexplored intersection of TS and neurodiversity by proposing a controlled experimental research design that combines subjective, physiological, and behavioral measures. The proposed design responds to three limitations in prior research: the limited integration of neurodiversity into TS research, the limited use of multimodal approaches in comparative studies across user groups, and the scarcity of controlled experimental studies in realistic digital work contexts.

By systematically comparing ND and NT individuals, the proposed design acknowledges that digital stress should not be treated as a homogeneous phenomenon. Combining self-reports with physiological and behavioral indicators supports a more differentiated understanding of stress responses, especially where perceived stress and objective indicators diverge.

Several limitations must be acknowledged. First, the proposed design is based on a laboratory setting, which may reduce ecological validity compared to real-world work environments [33]. Second, ND groups are inherently heterogeneous, and the selected conditions cannot represent the full spectrum of neurodiversity. Third, the proposed sample size and recruitment feasibility may constrain practical implementation.

Future research could extend this approach through field studies, larger and more differentiated samples, and the inclusion of contextual factors such as coping strategies, digital work experience, or workplace design.

## 5      Conclusion

Digitalization has fundamentally transformed modern work environments, while also introducing new forms of strain commonly described as TS. Within this research field, individual differences in how digital stressors are perceived and processed remain insufficiently understood, particularly with regard to neurodiversity. Existing TS research has largely focused on neurotypical populations and has rarely combined subjective, physiological, and behavioral perspectives within a single design.

To address these limitations, this paper proposes a controlled experimental research design that systematically compares ND and NT individuals under standardized digital stress conditions. By integrating structured and unstructured task scenarios with a multimodal measurement approach, the design enables a differentiated analysis of stress responses across multiple dimensions.

The key contribution of this work lies in explicitly linking neurodiversity to TS research and demonstrating the importance of considering heterogeneous cognitive and sensory processing patterns when studying digital stress. The proposed approach highlights that TS should not be treated as a uniform phenomenon, but rather as a complex interaction between technological demands and individual characteristics.

Future research can build on this design to conduct empirical studies, extend the approach to real-world work settings, and further explore how digital environments can be designed to support diverse user needs and promote more inclusive and cognitively adaptive workplaces.